\documentclass{article}
\pdfoutput=1  
\usepackage{amsmath}
\usepackage{amssymb,amsfonts}
\usepackage[all]{xy}
\usepackage{graphicx}

\numberwithin{equation}{section}
\numberwithin{table}{section}
\begin{document}

\thispagestyle{empty}


\vskip 3cm
\noindent
{\LARGE \bf Models for Modules}
\vskip .4cm
\noindent
{\Large  \bf \it
 The Story of {\cal O}   }
 \\
\vskip .1cm
\begin{center}
\linethickness{.06cm}
\line(1,0){447}
\end{center}
\vskip .5cm
\noindent
\noindent
{\large \bf Jan Troost}
\vskip 0.15cm
{\em \hskip -.05cm Laboratoire de Physique Th\'eorique\footnote{Unit\'e Mixte du CNRS et
    de l'Ecole Normale Sup\'erieure associ\'ee \`a l'Universit\'e Pierre et
    Marie Curie 6, UMR
    8549.}}
    \vskip 0cm
{\em \hskip -.05cm Ecole Normale Sup\'erieure}
 \vskip 0cm
{\em \hskip -.05cm 24 rue Lhomond, 75005 Paris, France}

\vskip 1cm

\vskip0cm

\noindent
{\sc Abstract:} 
We recall the structure of the indecomposable $sl(2)$ modules in the
Bernstein-Gelfand-Gelfand category ${\cal O}$. We show that all
these modules
can arise as quantized phase spaces of physical models.
In particular, we demonstrate in a
path integral discretization how a redefined action of the $sl(2)$
algebra over the complex numbers can glue finite dimensional and
infinite dimensional highest weight representations into
indecomposable wholes.   Furthermore, we
discuss how projective cover representations arise in the tensor
product of finite dimensional and Verma modules and give explicit tensor product decomposition
rules. The tensor product spaces can be realized in terms of
product path integrals.  Finally, we discuss
relations of our results to brane quantization and cohomological calculations
in string theory.

\vskip 1cm

\pagebreak

{
\tableofcontents }

\newpage

\section{Motivations}
The study of quantum physics mostly takes place in the arena of
unitary Hamiltonian time evolution in a Hilbert space of physical
states with positive norm. In many contexts though it turns out to be
useful to temporarily extend the arena of analysis to non-unitary
systems. This is true for example for open systems or for the
extension of unitary field theory Lagrangians to complexified
coupling constants.
It is also frequently useful to render the action of a symmetry
manifestly covariant, and this can also lead to the introduction of
null or negative norm states. The latter possibility arises in the
study of physical systems with a gauge symmetry, which includes many of
the basic theories of nature. Thus we may wish to temporarily study
non-unitary state spaces for physical systems.

In this paper, we study state spaces for systems with
$sl(2)$ symmetry, where $sl(2)$ denotes the three-dimensional Lie
algebra over the complex numbers. The state spaces will not necessarily
allow for an invariant positive norm. Our goal is to  generalize the
construction of a classical geometrical system whose path integral quantization
gives rise to spin
\cite{Nielsen:1987sa,Johnson:1988qm,Alekseev:1988vx}.  The
geometrical picture of finite dimensional $sl(2)$ representations was
inherited from the mathematical literature on the quantization of
coadjoint orbits of Lie groups \cite{Ko,S,W,Ki}. It applies to
representations that arise in the space of quadratically integrable
functions on the group. We broaden the path integral
treatment to include more general representations of $sl(2)$.  The
generalization can
be viewed as part of a program to render algebra geometric.

We will concentrate on $sl(2)$ modules with a finite number of highest
weight states\footnote{We take the mathematics convention here. In
  physics, we more often discuss representations with a finite
  number of lowest weight states.}. More precisely, we will
concentrate on the Bernstein-Gelfand-Gelfand category ${\cal O}$ of $sl(2)$
modules. These are modules that are finitely generated, that can be
decomposed into weight spaces, and that are locally finite with
respect to the action of raising operators (see
e.g. \cite{HumphreysBGG}).  These modules are classified. There are
examples amongst these modules that appear in various physical
contexts. These include the analytic continuation of
$SL(2,\mathbb{C})$ modules to $SL(2,\mathbb{R})$ modules (e.g. in
analytically continuing spaces of states on euclidean $AdS_3$ to
Lorentzian $AdS_3$ \cite{Satoh:2001bi}\cite{Maldacena:2001km}), and
open string representation spaces in the topological A-model on the
complexified sphere \cite{Gukov:2008ve}.

We were motivated for this study by the fact that if one considers
representations of the conformal group with lowest weight, and a set
of supersymmetry generators acting on those representations, one will
generate representations of the type that we analyze.  Equivalently,
they arise from the action of fermionic generators on representations of an
$AdS_d$ isometry group. We provide an opportunity
for understanding the appearance of projective modules and a non-diagonalizable action
of Casimirs in these contexts, using elementary physical models. These algebraic
phenomena are key in determining the space of physical states in Berkovits models of
superstrings on $AdS_d$ backgrounds with Ramond-Ramond flux \cite{Gotz:2006qp}\cite{Troost:2011fd}\cite{Gaberdiel:2011vf}.

\section{Representations}
\label{representations}
In this section we review the Bernstein-Gelfand-Gelfand (BGG) category ${\cal O}$ of $sl(2)$
modules. We refer the reader to the book \cite{HumphreysBGG} for a
detailed and pedagogical exposition.
\subsection{The BGG category ${\cal O}$}
  The $sl(2)$ algebra is generated by the three generators
$x,y,h$ that satisfy the commutation relations:
\begin{eqnarray}
  [x,y] =h \qquad [h,x]=2 x \qquad [h,y]=- 2 y. &&
\end{eqnarray}
We will concentrate on the modules of this algebra that can be
generated from a finite set of vectors. Moreover, the modules can be
decomposed into modules of given weight with respect to a Cartan
subalgebra, and each vector only generates a finite dimensional
subspace when we act on it with raising operators only. The
indecomposable modules inside this category of $sl(2)$ modules have
been classified (see e.g. \cite{HumphreysBGG}).  All other modules are
direct sums of these.  We review the indecomposable modules next.
\subsection{The indecomposable modules}
For any highest weight $\lambda \in \mathbb{C}$, we have a Verma
module $M(\lambda)$ which is generated by the free action of the
lowering operators.  It contains weights $\lambda, \lambda-2,
\lambda-4, \dots$.  When $\lambda$ is a positive integer, we have that the module
$M(\lambda)$ has a finite
dimensional simple module $L(\lambda)$ of dimension $\lambda+1$
as a quotient. In that case, we
have the short exact sequence $0 \rightarrow M(-\lambda-2) \rightarrow
M(\lambda) \rightarrow L(\lambda) \rightarrow 0$.

The category ${\cal O}$ 
also contains the modules $M^\vee(\lambda)$ dual to the
Verma modules.  The duality operation is such that it acts within the
BGG category ${\cal O}$.  The action of the $sl(2)$ generators on the
vector space dual to the original one is given by: $(h \cdot f)(v) =
f(h \cdot v)$, $(x \cdot f)(v) = f(y \cdot v)$ and $(y \cdot f) (v) =
f(x \cdot v)$ where the vector $v$ is in the original vector space and $f$ is a
map in the dual.  The modules $M^\vee(\lambda)$ have the property that
one can go from any state to the highest weight state by
ascent. This is dual to the property that in a Verma module
$M(\lambda)$, we can reach any state from the highest weight state by
descent. When $\lambda$ is a positive integer, we have the short
exact sequence $0 \rightarrow L(\lambda) \rightarrow M^\vee(\lambda)
\rightarrow M(-\lambda-2) \rightarrow 0$.

Finally, there are non-trivial projective modules $P(-\lambda-2)$ for
$\lambda$ a positive integer. These modules are indecomposable, have a
submodule $M(\lambda)$ and they fit into the short exact sequence $0
\rightarrow M(\lambda) \rightarrow P(-\lambda-2) \rightarrow
M(-\lambda-2) \rightarrow 0$. In fact, they also fit into the dual
short exact sequence $0 \rightarrow M(-\lambda-2) \rightarrow
P(-\lambda-2) \rightarrow M^\vee(\lambda) \rightarrow 0$. They are the
largest indecomposable modules that cover the Verma module
$M(-\lambda-2)$ (when $\lambda$ is a positive integer).

For the reader not familiar with these representations or their
description in terms of short exact sequences, it may prove useful to
study the modules in detail, using the explicit formulas given in
appendix \ref{detrep}. For more on the construction of the BGG
category, on simple, Verma and projective modules, we must refer to
\cite{HumphreysBGG} and references therein.

\section{Quantizations}
\label{quantizations}
In this section, we wish to realize the representations of section
\ref{representations} in
terms of phase spaces of models with a path integral formulation.
For each of the representations, we will specify the phase space, and
the expression of the generators of $sl(2)$ in the path integral
formalism, such that the quantum phase space becomes the desired
representation space. We hereby generalize the path integral
quantization of spin
\cite{Nielsen:1987sa}\cite{Johnson:1988qm}\cite{Alekseev:1988vx}.
\subsection{The finite dimensional representations}
\label{finpi}
\label{finite}
We first review the finite dimensional representations of $sl(2)$, and
their realization in a physical system.
\subsubsection{Orbit quantization}
 The finite dimensional representations $L(\lambda)$ are
unitary representations of the real form $su(2)$ of the algebra
$sl(2)$. They arise as Hilbert spaces from the geometric quantization
of the orbits of the $su(2)$ Lie algebra \cite{Ko,S,W,Ki}. A path
integral quantization of these orbits is known
\cite{Nielsen:1987sa}\cite{Johnson:1988qm}\cite{Alekseev:1988vx}, and
gives us the desired phase space quantization. For future
generalization, it is useful to review this construction and to add a
few details to the literature.

\begin{figure}
\centering
\includegraphics[width=6cm]{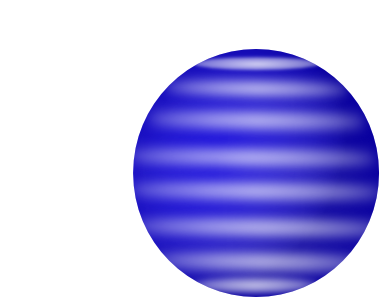}
\caption{The states in the Hilbert space correspond
to Bohr-Sommerfeld orbits at fixed angle. \label{finitereporbits}}
\end{figure}

The (co-adjoint) orbits of the Lie algebra $su(2)$ are two-spheres
(as in figure \ref{finitereporbits}).
They come equipped with a symplectic form $\Omega = (j+\frac{1}{2})
\sin \theta d \theta \wedge d \phi$ (where $\theta \in {[} 0, \pi {]}$
and $\phi \in {[} 0, 2 \pi {]}$). The total phase space volume is $2
\pi (2j+1)$, which gives rise to a state space of dimension
$\lambda+1=2j+1$. The phase space is generated by conjugation by group
elements from a (dual) Lie algebra vector of length
$j+\frac{1}{2}$. The action of the group on the space
is transitive, and thus gives rise
to an irreducible representation.  One considers a particle living on phase
space with an action given by $S= (j+\frac{1}{2}) \int d \tau \cos \theta
\dot{\phi}$. The Lagrangian is locally the integral of the symplectic
form.  A physical model for this action is an electron bound to a
sphere and only interacting electromagnetically with a magnetic
monopole located at the center of the sphere in a three-dimensional
space.  The prefactor is determined by the product of the chosen
electric and magnetic charges, which is quantized. The shift in the
spin is due to the metaplectic correction, or the improved Bohr-Sommerfeld
quantization condition.
One way to intuitively understand it is to note that the trivial
representation needs a single allowed Bohr-Sommerfeld orbit, embedded
in a sphere of total area $2 \pi$.
We have the  conserved charges\footnote{The link with the mathematics
notation is: $h=2 J^3$, $x = J^+$, $y=J^-$.}:
\begin{eqnarray}
J^3 &=& (j+\frac{1}{2}) \cos \theta
\nonumber \\
J^+ &=& (j+\frac{1}{2})\sin \theta e^{i \phi}
\nonumber \\
J^- &=&(j+\frac{1}{2})  \sin \theta e^{-i \phi}.
\end{eqnarray}
They Poisson commute into a $su(2)$ algebra.
We can reparameterize the charges using the variable $\eta = \cos \theta$
(where $\eta \in {[} -1, 1 {]}$):
\begin{eqnarray}
  J^3 &=& (j+\frac{1}{2}) \eta
  \nonumber \\
  J^+ &=& (j+\frac{1}{2})\sqrt{1-\eta^2} e^{i \phi}
  \nonumber \\
  J^- &=&(j+\frac{1}{2})  \sqrt{1-\eta^2} e^{-i \phi}.
\end{eqnarray}
The fundamental Poisson bracket reads $\{ \eta, \phi \} = (j+ 1/2)^{-1}$.
\subsubsection{The path integral}
To better understand the path integral quantization procedure
for the representations at hand, it will be useful to first perform a few
explicit calculations with the discretized path
integral for the finite dimensional representations.  See
e.g. \cite{Nielsen:1987sa}\cite{Johnson:1988qm}.
We first discuss the calculation of the expectation value of the angular
momentum and then that of the raising operator.
\subsubsection*{The angular momentum}
In the quantum theory, it will be useful to add a total
derivative term to the Lagrangian:
\begin{eqnarray}
S &=& 
 (j+\frac{1}{2}) \int d \tau \eta
\dot{\phi}
+
 \gamma \int d \tau
\dot{\phi}.
\end{eqnarray}
 We use the variables which are the
angle $\phi$ and the height parameter $\eta=\cos \theta$.  The discretized
path integral with insertion of the operator $J^3$ at a discrete
point in time labelled by $t_l$ is
then given by:
\begin{eqnarray}
Z (J^3(t_l)) &=& \lim_{N \rightarrow \infty}
\frac{1}{(2 \pi)^N}
\prod\int_{-1}^{+1} d \eta_{k=1,\dots,N} \int_{0}^{2 \pi}
d \phi_{k=1,\dots,N-1}
\nonumber \\
&  &
e^{i \sum_{k=1}^N (j+\frac{1}{2}) \eta_k (\phi_k-\phi_{k-1})
+ i\gamma (\phi_k - \phi_{k-1})}
 (j+\frac{1}{2}) 
\eta_l.
\end{eqnarray}
We integrate over all discrete momenta $\eta_k$, and over the
intermediate positions $\phi_k$, while specifying boundary conditions
$\phi_{0}$ and $\phi_N$ on the path integral. We use the Haar measure.
These finite dimensional integrals are straightforwardly computed:
\begin{eqnarray}
Z (J^3(t_l))&=&  \frac{1}{2 \pi} \int_{-1}^{+1}
d \eta e^{i (j+\frac{1}{2}) \eta (\phi_N- \phi_0) + i \gamma(\phi_N-\phi_0)}
 (j+\frac{1}{2}) \eta.
\end{eqnarray}
To render the path integral periodic in the angular variable $\phi_N$,
we sum over all angles $\phi_N$ that differ by $2 \pi n$
where $n \in \mathbb{Z}$. The term in the action proportional
to the angle $\gamma$ associates a phase $e^{2 \pi i \gamma n}$
to the winding sector labelled by $n$.  We obtain:
\begin{eqnarray}
Z (J^3(t_l))&=&
 \sum_{k \in \mathbb{Z}} \int_{-1}^{+1}
d \eta 
\delta (  (j+\frac{1}{2})\eta + \gamma -k)
e^{i  (j+\frac{1}{2})\eta (\phi_N- \phi_0) + i \gamma(\phi_N-\phi_0)}
 (j+\frac{1}{2}) \eta.
\label{angularmomentum}
\end{eqnarray}
If we Fourier transform the initial condition $\phi_0$ and the final
condition $\phi_N$ to dual integers $p_0$ and $p_N$, we find that
$(j+\frac{1}{2}) \eta+\gamma$ is equal to both. Indeed, the angular
momentum is conserved over the course of the time evolution. We note
that the angular momentum $(j+\frac{1}{2})\eta$ is an integer minus
$\gamma$. We can therefore postulate that for unitary finite
dimensional representations of $su(2)$ with integer spin $j$ the
parameter $\gamma$ is zero, while for half-integer spin $j$, the
parameter $\gamma$ is half-integer.  This presupposes that the
expression for the angular momentum $J^3$ is unmodified in the quantum
theory. (See
e.g. \cite{Nielsen:1987sa}\cite{Johnson:1988qm}\cite{Gukov:2008ve}\cite{Troost:2003ge}
for discussions of the parameter $\gamma$.)
\subsubsection*{The raising operator}
We turn to the study of the action of the raising operator in the
quantum theory. It is important to know when the raising operator
annihilates a state, and therefore to resolve the quantum ambiguities in 
its definition, and in the choice of regularization.
The path integral with an insertion of  the raising
operator $J^+$ at time $t_l$
can be discretized with a mid-point prescription as follows: 
\begin{eqnarray}
Z (J^+(t_l)) &=& \lim_{N \rightarrow \infty}
\frac{1}{(2 \pi)^N}
\prod \int_{-1}^{+1}d \eta_{k=1,\dots,N} 
\int_0^{2 \pi}
d \phi_{k=1,\dots,N-1}
\nonumber \\
&  &
e^{i \sum_{k=1}^N (j+1/2) \eta_k (\phi_k-\phi_{k-1})
+ \gamma (\phi_k - \phi_{k-1})}
 (j+\frac{1}{2}) 
\sqrt{1-\eta_l^2} e^{\frac{i}{2} (\phi_l+\phi_{l-1})}
\end{eqnarray}
with initial value $\phi_0$ and final condition $\phi_N$.  The $N-1$
integrals over the variables $\phi_k$ give rise to as many
delta-functions. This leads to the constraints
$(j+\frac{1}{2}) \eta_l = (j+\frac{1}{2})\eta_{1,2,\dots,l-1}+1/2$ and
$(j+\frac{1}{2}) \eta_{l+1,l+2,\dots,N} = (j+\frac{1}{2}) \eta_1
+1$. Thus, the path integral reduces to:
\begin{eqnarray}
Z (J^+(t_l)) &=& \frac{1}{2 \pi} \int_{-1}^{+1}
d \eta e^{i(j+\frac{1}{2}) \eta (\phi_N- \phi_0) + i \gamma(\phi_N-\phi_0)}
e^{i \phi_N}
 (j+\frac{1}{2}) \sqrt{1 - (\eta+\frac{1}{2j+1})^2}.
\end{eqnarray} 
To get a periodic result, we again sum over final conditions
$n \in \mathbb{Z}$ where $\phi_N(n)
= \phi_N + 2 \pi n$. This gives rise to:
\begin{eqnarray}
Z (J^+(t_l)) &=& \sum_{k \in \mathbb{Z}} \int_{-1}^{+1}
d \eta 
\delta ( (j+ \frac{1}{2}) \eta + \gamma -k)
e^{i(j+\frac{1}{2}) \eta (\phi_N- \phi_0) + i \gamma(\phi_N-\phi_0)}
e^{i \phi_N}
\sqrt{ (j+\frac{1}{2})^2 - ((j+\frac{1}{2}) \eta+\frac{1}{2})^2}.
\nonumber
\end{eqnarray}
We again Fourier transform the boundary
conditions $\phi_{0,N}$  on the path integral:
\begin{eqnarray}
Z (J^+(t_l)) 
 &=& \sum_{k \in \mathbb{Z}} \int_{-1}^{+1}
d \eta 
\delta (  (j+\frac{1}{2}) \eta + \gamma  -k)
\delta(p_0 - p_N +1)
\delta (p_N -  (j+\frac{1}{2}) \eta -  \gamma-1)
\nonumber \\
& &
 \sqrt{ (j+1/2)^2 - (m+1/2)^2}
\nonumber
\end{eqnarray}
The result is non-zero only when the initial and final angular
momentum $m=(j+1/2)\eta$ differ by one.  Again, at integer spin $j$,
the parameter $\gamma$ is zero, while at half-integer spin, it is
half-integer.  {From} standard $su(2)$ representation theory, the
expected coefficient for the action of the raising operator is:
\begin{eqnarray}
\sqrt{j (j+1) - m(m+1)}
&=&
\sqrt{ (j+1/2)^2 - (m+1/2)^2}.
\label{coefrais}
\end{eqnarray}
We see that this agrees with our path integral result, with a
definition for the raising operator $J^+$ which coincides with the
classical operator (when using the mid-point prescription). {From} now on,
we will assume that our quantization procedure agrees with
the intuition that the classical vanishing of the raising operator indicates
the appearance of a maximal vector in the quantum state space (with quantum 
eigenvalue $m$ just below the classical value).
   
\subsection{The highest weight discrete representations} 
\label{discpi}
\begin{figure}
 \centering
 \includegraphics[angle=180,width=6cm]{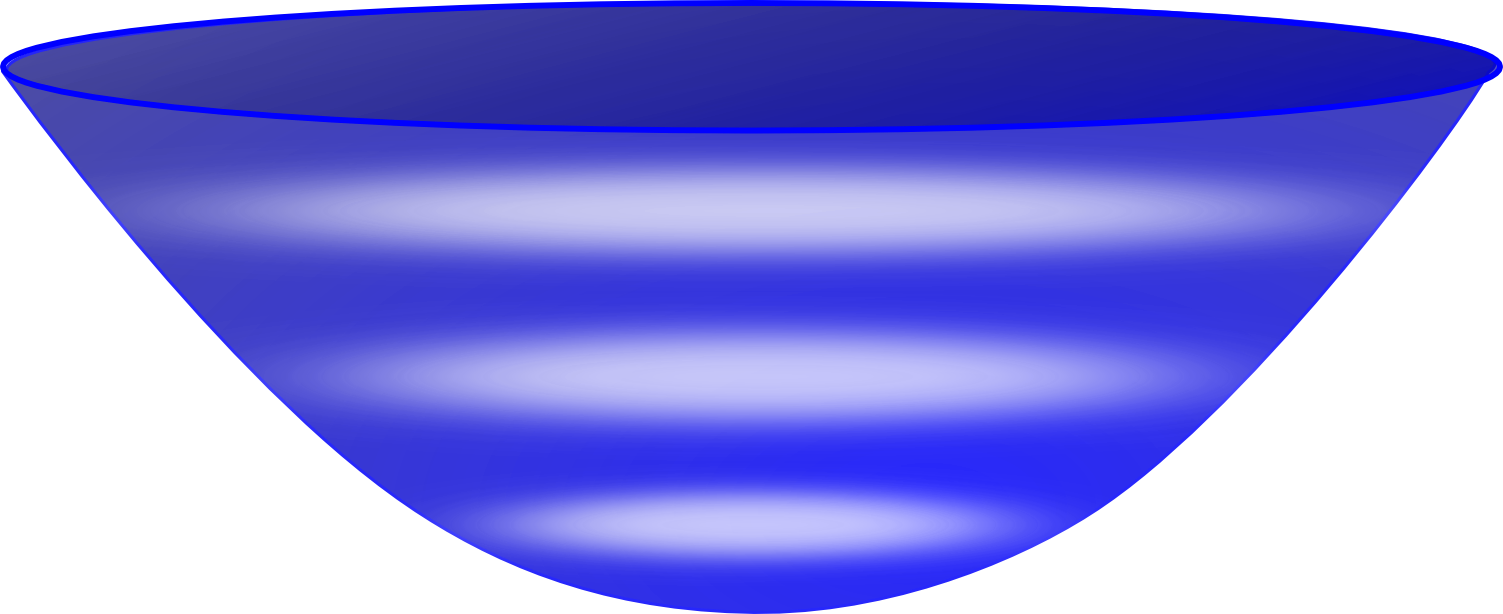}
 \caption{Bohr-Sommerfeld orbits for a highest weight discrete representation.
\label{discrete}
}
 \end{figure}

A path integral description of the quantum system with a Hilbert space
consisting of a highest weight discrete representation with real highest
weight lower than $-1/2$ can also be constructed through the
coadjoint orbit method applied to $sl(2,\mathbb{R})$ (see
e.g. \cite{Witten:1987ty}\cite{Troost:2003ge} and figure \ref{discrete}).  The associated action
is of the form:
\begin{eqnarray}
S &=& -(j+1/2) \int d \tau \cosh \rho \dot{\phi} + \gamma \int d \tau
\dot{\phi}
\end{eqnarray}
where $j>-1/2$.
Instead of integrating the coordinate $\eta$ from $-1$ to $+1$, we
integrate $\eta=-\cosh \rho$ from $- \infty$ to $-1$.  For a given
positive $j$ and a canonical choice of the parameter $\gamma$, the
first eigenvalue of the angular momentum operator will be
$-(j+1)$. Thus, the Hilbert space will be the highest weight
representation of $sl(2)$ which is sometimes denoted as
$M(-2-2j)=D^{-}_{j+1}$. Before we generalize this construction, we draw
a few lessons.

\subsection*{Lessons}
It should be clear that we can treat the finite dimensional and
semi-infinite examples uniformly, by using the variables $\phi$ and
$\eta$ to parameterize the phase space.  The Poisson bracket of the
position and momentum space variables is:
\begin{eqnarray}
\{ \eta, \phi \} &=& \frac{1}{j+\frac{1}{2}}
\end{eqnarray}
in both the above systems. We have a line segment or half-line
with coordinate $\eta$, and a circle bundle over it, with fiber
parameterized by $\phi$.
The size of the circle fibered over the line segment
 depends on the point $\eta$
like $\sqrt{1- \eta^2}$ 
on the interval $\eta \in {[} -1,+1 {]}$, while it is equal to 
$\sqrt{\eta^2-1}$ for points with parameter $\eta$ smaller than $-1$.
In these variables, the $sl(2)$ algebra takes the form (up to a factor
of $\pm i $):
\begin{eqnarray}
\label{originalcharges}
J^3 &=& (j+\frac{1}{2}) \eta
\nonumber\\
J^\pm &=& (j+\frac{1}{2}) \sqrt{|1-\eta^2|} e^{\pm i \phi}. 
\end{eqnarray}

\subsection{The Verma modules}
\label{verma}
\begin{figure}
 \centering
 \includegraphics[width=6cm]{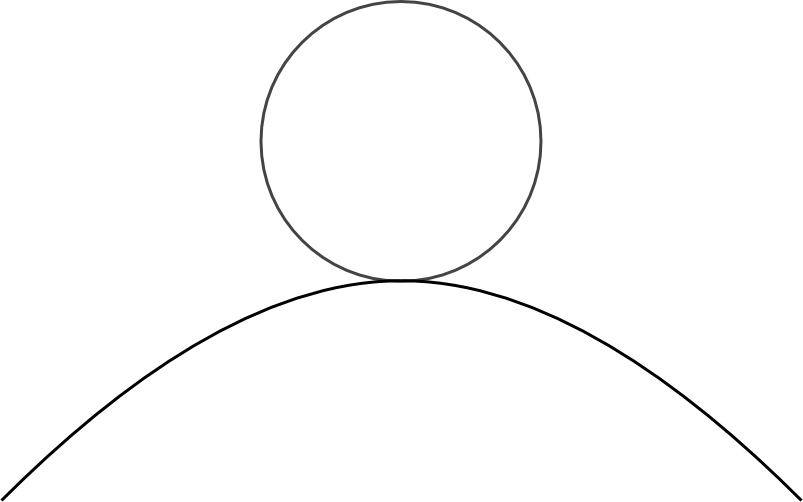}
 \caption{The sphere and the hyperboloid with the same value for the quadratic
Casimir touch. \label{circlehyperboloid}}
 \end{figure}

We have reviewed the geometric path integral for spin for
finite dimensional representations and certain highest weight
representations. Our goal is to construct path integral formulations
for more general representations. We first consider a positive integer
$\lambda$ and the highest weight representations $M(\lambda)$. {From}
the weight spaces of the Verma module $M(\lambda)$ it is clear that we
can identify its phase space as consisting of the spherical phase
space corresponding to a finite dimensional representation
$L(\lambda)$, combined with a hyperboloidal phase space of a Verma
module $M(-\lambda-2)$ (see figures \ref{finitereporbits}, \ref{discrete} and \ref{circlehyperboloid}).
 If we quantize
the unified phase space as we did previously, with the action of the lowering
and raising operators that we found before, then the space of
states will take the direct sum form. We must make a modification
such as to glue the two parts of the phase space into a Verma module
$M(\lambda)$. In particular, we must avoid the annihilation of the lowering
operator at the bottom of the sphere.

\begin{figure}
 \centering
 \includegraphics[width=6cm]{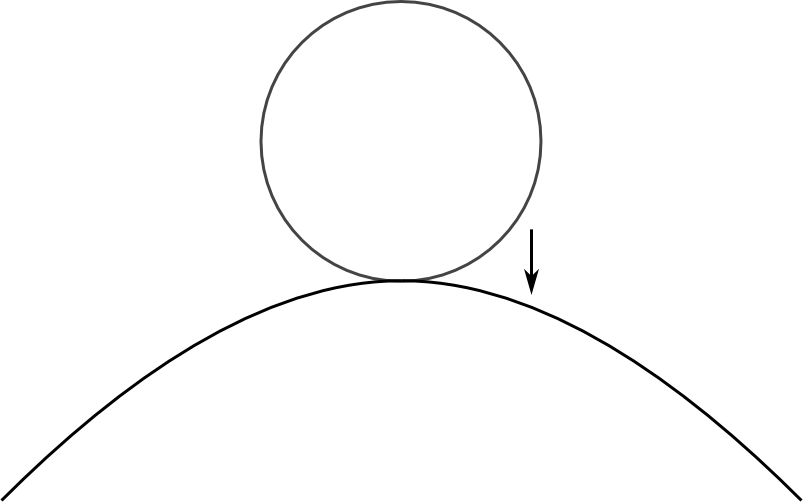}
 \caption{The lowering operator connects the sphere and the hyperboloid in a Verma module
with positive highest weight. \label{down}}
 \end{figure}

To understand the gluing procedure, we first analyze a classical
counterpart. If we suppose that the angular momentum generator still
takes the form $J^3  \propto \eta$, then we have for any raising and lowering
operator of the form $J^\pm = f^\pm(\eta) e^{\pm i \phi}$ that the
Poisson bracket $\{ J^3, J^\pm \} \propto  \pm i J^\pm$ is satisfied.
Let's compute the bracket of two such charges:
\begin{eqnarray}
\{ J^+, J^- \} & \propto & 
i ( f^- \partial_\eta f^+ + f^+ \partial_\eta f^-). 
\end{eqnarray}
For given solutions $f^\pm$ to the commutation relation
$\{ J^+,J^- \} \propto  2 i J^3$, we have that any variation in which
we multiply both by inverse functions $f^\pm \rightarrow f^\pm g^{\pm 1}$
will still satisfy the
same Poisson bracket relations. Thus, to
glue two representations, we rescale away the zero associated to the size of the circle
fiber in
either the raising or the lowering operator. 
In practice, we  
define the $sl(2)$ operators as:
\begin{eqnarray}
J^3 &=& \frac{\lambda+1}{2} \eta
\nonumber \\
J^+ &=& \frac{\lambda+1}{2}(1-\eta^2) e^{i \phi}
\nonumber \\
J^- &=& \frac{\lambda+1}{2}e^{- i \phi}.
\label{gluedcharges}
\end{eqnarray}
We consider the path integral which consists of the two path
integrals we described in subsections \ref{finpi} and
\ref{discpi}, but  now we integrate over the full region of the
variable $\eta$. It should be clear that, first of all, the angular
momentum spectrum will be the spectrum of the direct sum of the finite
representation $L(\lambda)$ and the Verma module $M(-\lambda-2)$,
since the range of integration of $\eta$ now covers both regions that
previously gave rise to discretized angular momenta. The symmetry algebra is still
realized.  And, crucially, the lowering operator $J^-$ will not longer
annihilate the state sitting at the bottom of the spherical part of
the phase space (see figure \ref{down}).
 We thus necessarily
generate the desired representation $M(\lambda)$.  We can indeed check that
the raising
operator will still annihilate the top state in the representation,
as well as the top state in the submodule $M(-\lambda-2)$. This follows from
formula (\ref{coefrais}), and the fact that the raising operator 
in equation (\ref{gluedcharges})
gives rise to the coefficient (\ref{coefrais}) squared. This achieves the desired
goal of a path integral realization of the Verma module $M(\lambda)$.

\subsection*{An Invariant}
We note that when we consider the $sl(2)$ algebra as realized in the
direct sum representation, then the step function $\theta(\eta+1)$ is
an invariant.  The step function Poisson-commutes with the charges in
the direct sum (as in equation (\ref{originalcharges})) but not  with
the charges realizing the Verma module $M(\lambda)$ (as in equation
(\ref{gluedcharges})).  This accords
with intuition since the function assigns one value to one direct
summand, and another to the second summand. This is not an invariant
if we connect the two summands through the action of the $sl(2)$
algebra.

\subsection*{Analytic continuation}
Finally we note that the operators (\ref{gluedcharges}) as well as the symplectic
structure are analytic in $\lambda$ (except at $\lambda=-1$). They
allow for analytic continuation of our construction of Verma modules
$M(
\lambda)$ to all $\lambda \in \mathbb{C}
\backslash \{ -1 \}$. At $\lambda=-1$, the fixed point of the shifted Weyl 
reflection $\lambda \leftrightarrow -\lambda -2$,  the symplectic structure becomes singular.
We treat this special case next.

\subsection{The mock discrete representation $M(-1)=D^-_{1/2}$}
\label{mock}
We can obtain the exceptional Verma module $M(-1)=D^-_{\frac{1}{2}}$
by quantizing the action:
\begin{eqnarray}
S &=& -\int d \tau e^{\rho} \dot{\phi}+ \gamma \int d \tau \dot{\phi}.
\end{eqnarray}
We have the variable $\eta = -e^{\rho} \in {]} -\infty, 0 ]$.
The generators of $sl(2)$ take the form:
\begin{eqnarray}
J^3 &=& \eta
\nonumber \\
J^+ &=& \eta^2 e^{i \phi}
\nonumber \\
J^- &=&  -e^{- i \phi}.
\end{eqnarray}
Quantizing as we did for finite or discrete highest weight
representations (with $\gamma=1/2$) gives rise to angular momenta $m
\in \{ -1/2,-3/2,\dots \}$ (because of the choice of parameter
$\gamma$ and the range of integration of the coordinate $\eta$ --
compare to equation (\ref{angularmomentum})). The expression for the
coefficient of the raising operator will now be $(m+\frac{1}{2})^2$ --
compare to equation (\ref{coefrais}) -- which has a double zero. The
lowering operator will have coefficient $-1$ in terms of the same
basis of states.  This indeed gives rise to the Verma module $M(-1)$,
or equivalently, the mock discrete representation $D^-_{1/2}$.  This
is an exceptional case since it does not arise from coadjoint
orbit quantization. It can be obtained by a limiting procedure
(in which one rescales both the variable $\eta$ as well as the charges $J^\pm$)
from the Verma modules treated in subsection \ref{verma}.
The representation plays a special role due to the fact that its
highest weight is self-mirror under Weyl reflection (shifted by half
the sum of the positive roots).

\subsection{The dual Verma modules}
It is an easy exercise to show that the dual Verma modules
$M^\vee(\lambda)$ as well are amenable to path integral
quantization. It is sufficient to shift the zero 
from the raising to the lowering operator in the charges in equation
(\ref{gluedcharges}).

\subsection{The projective modules}
We would like to extend the construction above to cover path integral
realizations of the (non-trivial) projective  modules of the BGG category ${\cal O}$.  For
the projective representations $P(-\lambda-2)$, where $\lambda$ is a
positive integer, one might expect a (partially) double-sheeted phase space
corresponding to weight spaces of dimension two.
However, within the projective representation, we would need to be
able to move from one such sheet to the other.  We need to
quantize a geometry that  incorporates this feature in its classical guise. 
Before we discuss this geometric realization, it will be
useful to first study how projective representations arise in tensor
product modules.

\section{Multiplications}
In this section, we study how various modules in the BGG category
${\cal O}$ arise from more familiar modules, through the operation of
taking tensor products. We will give explicit decomposition rules for
some tensor product representations. Tensor products between finite and infinite
dimensional modules have been studied in the mathematics literature
(see e.g. \cite{KoTens,Z,BG}) but explicit formulas for decompositions 
of tensor products are hard to find. The decomposition
rules can be obtained as in \cite{BSIII} from applying special
projective functors to diagram algebras\footnote{We would like to
  thank Catharina Stroppel for pointing this out.} -- here we follow a
more pedestrian approach.

\subsection{Tensoring finite dimensional modules}
We start out with finite dimensional modules. Those all arise from
taking consecutive tensor products of the two-dimensional
representation with itself. We have the standard decomposition
formula for spin:
\begin{eqnarray}
L(\lambda_1) \otimes L(\lambda_2)
&=& L(|\lambda_1-\lambda_2|) \oplus L(|\lambda_1-\lambda_2|+2)
\oplus \dots \oplus L(\lambda_1+\lambda_2).
\end{eqnarray}
\subsection{Tensoring finite dimensional modules with Verma modules I}
The next operation that we wish to study is the tensor product of
finite dimensional representations with other indecomposable
representations in the category ${\cal O}$. The result will always lie
in the category ${\cal O}$.  Let's study the tensor product of a
finite dimensional representation $L(\lambda_2)$ and an infinite
dimensional highest weight representation, the Verma module
$M(\lambda_1)$. Since the tensor
product operation is associative, and since all finite dimensional
representations can be obtained by taking tensor products of the
two-dimensional representation with itself, we can restrict our
attention to the tensor product of the two-dimensional representation
with a generic Verma module $M(\lambda_1)$. We can decompose the
result in terms of direct summands characterized by their character. A
central element of the universal enveloping algebra will act on each
such direct summand as a scalar character plus a nilpotent operator (see e.g.
\cite{HumphreysBGG}.

We moreover know \cite{HumphreysBGG} that the module $M(\lambda_1)
\otimes L(1)$ permits a standard filtration (i.e. a filtration in
terms of Verma modules) where the Verma modules arising as quotients
are $M(\lambda_1-1)$ and $M(\lambda_1+1)$.  The weights $\lambda_1 \pm
1$ are only linked when $\lambda_1=-1$ (i.e. these Verma modules can
only combine into a non-direct sum representation when this condition
is satisfied).  In all other cases, the tensor product will be a
direct sum of the two Verma modules. Note that for $\lambda_1=-1$ we
have that $M(\lambda_1)$ is projective. Therefore, the tensor product
with the finite dimensional representation will also be
projective. The only possibility for the result of the tensor product
is then the projective cover $P(-2)$.  We summarize:
\begin{eqnarray}
M(-1) \otimes L(1) &=& P(-2)
\nonumber \\
M(\lambda_1) \otimes L(1)
&=& M(\lambda_1+1) \oplus M(\lambda_1-1) \qquad \mbox{otherwise}.
\end{eqnarray}
We can summarize this in words by saying that any Verma modules
appearing in the standard filtration of these tensor products that can
team up will.  Since we used some abstract nonsense to arrive at
these results, it may be good to also demonstrate the non-trivial
tensor product hands-on.  In appendix \ref{ETP}, we demonstrate
through explicit calculation that these tensor product formulas hold.

It is clear now that to recursively compute tensor products of Verma
modules with finite dimensional representations, we need to analyze
the tensor product of projectives with finite dimensional modules
first.
\subsection{Tensoring finite dimensional with projective modules}
We analyze the tensor product of the projective covers $P(-n)$ where
$-n \in \{-2,-3,\dots \}$ with finite dimensional representations. 
\subsubsection*{The basis for induction}
 We
 start out with the calculation of the tensor product $P(-n)
\otimes L(1)$.  The result is necessarily projective, and therefore
permits a filtration with Verma modules. By an analysis of the weight
spaces, one sees that the standard filtration contains the Verma
modules $M(-n+1), M(-n-1),M(n-3),M(n-1)$.  These are linked two by
two, namely $M(-n+1),M(n-3)$ and $M(-n-1),M(n-1)$.  The Verma module
$M(-n-1)$ must appear as a factor in its projective cover $P(-n-1)$,
which is a direct summand. Generically, this is also true for the
Verma module $M(-n+1)$, which will appear in the summand
$P(-n+1)$. The only exception that can occur is when $-n=-2$. In that
case, we have that the Verma module $M(-1)$ is projective by
itself. It then appears with multiplicity $2$ in the decomposition.
We summarize:
\begin{eqnarray}
P(-2) \otimes L(1) &=& P(-3) \oplus 2 M(-1)
\nonumber \\
P(-n) \otimes L(1) &=& P(-n-1) \oplus P(-n+1)
 \qquad \mbox{otherwise}.
\end{eqnarray}
Again, any Verma modules appearing in the standard filtration of the
product that can link up must.
\subsubsection*{Notation}
The following notation will be useful. For all $x \in \mathbb{Z}$,
we define
\begin{eqnarray}
{\cal P}(x) &=& P(-|x+1|-1)
\end{eqnarray}
where it is understood that the notation $P(-1)$ stands for the module
$M(-1)$ with multiplicity two, $P(-1)=2M(-1)$. (This is an abusive
notation that will be handy.) We can then compute the tensor product:
\begin{eqnarray}
{\cal P}(x) \otimes L(1)
&=& {\cal P} (x+1) \oplus {\cal P}(x-1).
\end{eqnarray}
This can be checked on a case by case basis. The formula find its origin
in the fact that Verma modules related by shifted Weyl reflection pair
up in projective covers.
\subsubsection*{The induction step}
We are now ready to prove the tensor product decomposition formula
of projective modules $P(-n)$ with the finite dimensional
representations. 
We claim the tensor product formula:
\begin{eqnarray}
P(-n) \otimes L(\lambda_2)
&=& 
{\cal P}(-n-\lambda_2)\oplus {\cal P}(-n-\lambda_2+2) \oplus \dots
\oplus {\cal P}(-n+\lambda_2-2)
\oplus
 {\cal P}(-n+\lambda_2). 
\label{projfin}
\end{eqnarray}
The basis for the induction was proven,
namely, for arbitrary $n=-2,-3,\dots$ and for $\lambda_2=1$.
Suppose now the formula holds for given $n$ and $\lambda_2$ (as well as 
$\lambda_2-1$). Then we tensor
multiply $L(1)$ on both sides of the decomposition formula.
We use the decomposition $L(\lambda_2) \otimes L(1) 
= L(\lambda_2-1) \oplus L(\lambda_2+1)$ as well as the fact
that the formula holds for arbitrary $n$ and $\lambda_2-1$. We then
subtract the result for $P(-n) \otimes L(\lambda_2-1)$ from both
sides. That proves the induction step, and therefore the tensor product
decomposition formula (\ref{projfin}).
\subsection{Tensoring finite dimensional modules with Verma modules II}
\begin{figure}
 \centering
 \includegraphics[width=12cm]{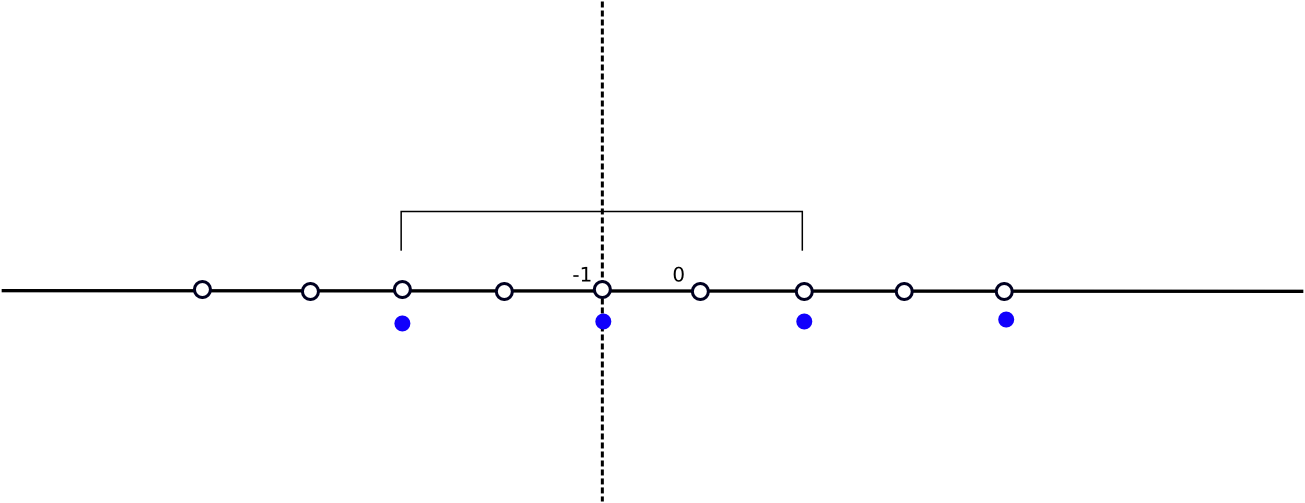}
 \caption{A Verma module tensored with a finite dimensional representation gives rise to a module
with a standard filtration in which various Verma modules appear (with multiplicity one). Those
Verma module that can pair up into a projective module, will. This figure
illustrates the standard filtration (just below the horizontal weight axis) 
and the pairing for the tensor product
$M(0) \otimes L(3) = M(-1) \oplus P(-3) \oplus M(3)$. \label{weightspace} }
 \end{figure}

We turn to applying this insight to the tensor product of
Verma modules with finite dimensional modules. Again we prove the
decomposition formula by induction.
For the induction hypothesis, we use the maxim that anything that can
pair up, will (see figure \ref{weightspace}). Consider the tensor product $M(\lambda_1) \otimes
L(\lambda_2)$. It has a Verma module filtration with modules
$M(\lambda_1-\lambda_2), M(\lambda_1-\lambda_2+2), \dots ,
M(\lambda_1+\lambda_2)$ centered around $\lambda_1$. In order for a possibility for pairing to
occur, we must have that either $0$ and $-2$, or $+1$ and $-3$ is
amongst the highest weights of these Verma modules. Otherwise, we have
a direct sum of Verma modules. When $0,2$ or $+1,-3$
are amongst these weights, 
the result
will  depend on whether $\lambda_1$ is smaller than $-1$ or
bigger than $-1$. We propose:
\begin{eqnarray}
M(\lambda_1) \otimes L(\lambda_2) &=& 
M(\lambda_1-\lambda_2) \oplus M(\lambda_1-\lambda_2-2) \oplus \dots \oplus
M(\lambda_1+\lambda_2)
\nonumber \\
& &\mbox{when}  \,\, \lambda_1-\lambda_2 \ge -1 \,\, 
\mbox{or} \,\,   \lambda_1+\lambda_2 \le -1, \mbox{and otherwise:}
\nonumber \\
M(\lambda_1) \otimes L(\lambda_2) &=& P(-2) \oplus P(-4)
\oplus P(-\lambda_1-\lambda_2-2)
\oplus M(-\lambda_1-\lambda_2-4) \oplus \dots \oplus
M(\lambda_1-\lambda_2)
\nonumber \\
& & 
 \mbox{for} \,\, \lambda_1 \le -1 \,\, \mbox{and} \,\, \lambda_1-\lambda_2 \,\, \mbox{even}
\nonumber \\
M(\lambda_1) \otimes L(\lambda_2) &=& M(-1) \oplus P(-3) \oplus P(-5)
\oplus P(-\lambda_1-\lambda_2-2)
\oplus M(-\lambda_1-\lambda_2-4) \oplus \dots \oplus
M(\lambda_1-\lambda_2)
\nonumber \\
& &
\mbox{for} \,\, \lambda_1 \le -1 \,\,  \mbox{and} \,\,  \lambda_1-\lambda_2 \,\, \mbox{odd}
\nonumber \\
M(\lambda_1) \otimes L(\lambda_2) &=& P(-2) \oplus P(-4)
\oplus P(\lambda_1-\lambda_2)
\oplus M(\lambda_2-\lambda_1) \oplus \dots \oplus
M(\lambda_1+\lambda_2)
\nonumber \\
& &
\mbox{for} \,\,  \lambda_1 \ge -1 \,\,  \mbox{and} \,\, \lambda_1-\lambda_2 \,\,  \mbox{even}
\nonumber \\
M(\lambda_1) \otimes L(\lambda_2) &=&M(-1) \oplus P(-3) \oplus P(-5)
\oplus P(\lambda_1-\lambda_2)
\oplus M(\lambda_2-\lambda_1) \oplus \dots \oplus
M(\lambda_1+\lambda_2)
\nonumber \\
& &
\mbox{for}\,\,  \lambda_1 \ge -1 \,\,  \mbox{and} \,\,  \lambda_1-\lambda_2 \,\,  \mbox{odd}
\label{finVerm}
\end{eqnarray}
To prove this we use induction on $\lambda_2$. Suppose the induction
step is true for $\lambda_2$ (and smaller weights).
Let's continue the proof in a particular case. The
other cases are proved analogously.
Suppose that $\lambda_1 \le -1$ and $\lambda_1-\lambda_2$ even,
and that $P(-2)$ is amongst the direct summands of $M(\lambda_1)
\otimes L(\lambda_2)$. We  have that the  decomposition
after tensoring in $L(1)$ on the right hand side of the induction
hypothesis (namely the second line of (\ref{finVerm})) gives rise to the representations:
\begin{eqnarray}
& & 2 M(-1) \oplus 2 P(-3) \oplus \dots
2 P(-\lambda_1-\lambda_2-1) 
\oplus P(-\lambda_1- \lambda_2 -3)
\nonumber \\
& & 
\oplus M(-\lambda_1-\lambda_2-3) \oplus
2 M(-\lambda_1-\lambda_2-5) \oplus \dots
\oplus 2 M(\lambda_1-\lambda_2+1) 
\oplus M(\lambda_1-\lambda_2-1).
\end{eqnarray}
 We then use the induction hypothesis (namely the third line of (\ref{finVerm}))
 on the tensor product $M(\lambda_1) \otimes L(\lambda_2-1)$
to take out the direct summands:
\begin{eqnarray}
M(-1) \oplus P(-3) \oplus \dots \oplus P(-\lambda_1-\lambda_2-1)
\oplus M(-\lambda_1-\lambda_2-3) \oplus \dots \oplus M(\lambda_1-\lambda_2+1)
\end{eqnarray}
and we're left with:
\begin{eqnarray}
M(\lambda_1) \otimes L(\lambda_2+1) &=& M(-1) \oplus
\oplus \dots 
\oplus P(-\lambda_1-\lambda_2-3)
\oplus M(-\lambda_1-\lambda_2-5) \oplus \dots \oplus M(\lambda_1-\lambda_2-1)
\nonumber
\end{eqnarray}
which proves the next step in the induction procedure.

\subsection{Tensoring finite dimensional with dual Verma modules}
The tensor product of finite dimensional modules with dual Verma
modules is dual to the tensor product of finite dimensional modules
with ordinary Verma modules. Thus, to obtain the tensor product
decomposition formulas, we apply duality to the decomposition formulas we obtained
before. Using that both projective and finite dimensional modules are
self-dual, we get the desired result.

\subsection{The path integral representation for projective modules}
\begin{figure}
 \centering
 \includegraphics[width=6cm]{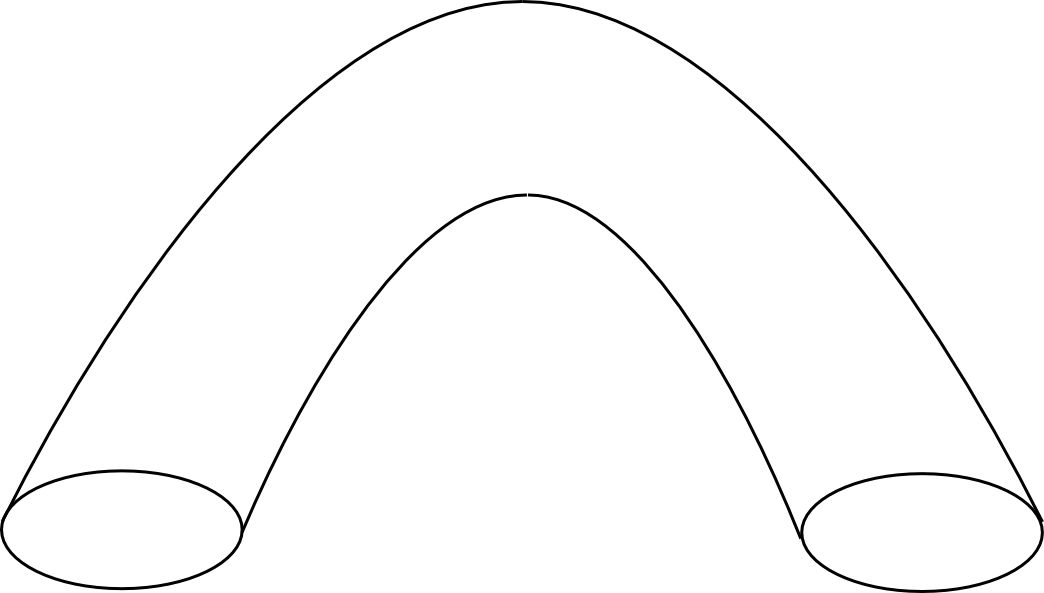}
 \caption{A sketch of the geometry of the projective
   representation $P(-2)$ arising from the tensor product of a highest
   weight representation and a two-dimensional representation with
   spherical phase space.  \label{projective} }
 \end{figure}

Since we have a path integral realization of the Verma modules and the
finite dimensional modules in the category, it should be clear that
their tensor product representations can be realized by taking
products of the corresponding path integrals. Therefore, we now have
in hand path integral representations for all (non-trivial) projective
modules $P(-n)$ as well. For concreteness, we sketch the simplest
example, in parallel to the discussion of the tensor product
decomposition formulas. We start from the path integral representation
of the Verma module $M(-1)$ (as in subsection \ref{mock}) tensored
with the path integral corresponding to the two-dimensional module $L(1)$
(described in subsection \ref{finite}). It is clear that the latter
module will make sure that we can have two-dimensional weight spaces
that are non-trivially interconnected in the quantum theory (see figure \ref{projective}).
We now integrate over
two variables $\eta_1 \in ] - \infty , 0 ]$ and $\eta_2 \in
[-1,1]$ as well as  two angle variables $\phi_{1,2} \in [0, 2
\pi]$. The expressions for the charges are given by the sums:
\begin{eqnarray}
J^3 &=& \eta_1 + \eta_2
\nonumber \\
J^+ &=& \eta_1^2 e^{i \phi_1} + \sqrt{1-\eta_2^2} e^{i \phi_2}
\nonumber \\
J^- &=& - e^{-i \phi_1} + \sqrt{1-\eta_2^2} e^{-i \phi_2}.
\end{eqnarray}
We perform the path integral quantization with the action:
\begin{eqnarray}
S &=& \int d \tau \left( (\eta_1+ \frac{1}{2}) \dot{\phi_1}
+ (\eta_2+\frac{1}{2}) \dot{\phi_2} \right),
\end{eqnarray}
and due to the analysis of the factor path integrals given in 
subsections \ref{finite} and \ref{mock}, the resulting phase space will
correspond to the tensor product $M(-1) \otimes L(1) $ which we have
proven to be equal to the projective representation $P(-2)$. The other projective representations
are found as direct summands in other tensor products that can also be obtained by product
path integral quantizations. One can for instance isolate a given projective summand $P(-n)$
appearing in the tensor product of Verma modules with finite dimensional modules by projecting on 
a particular value of the quadratic Casimir squared.

\subsubsection*{Summary}
For each module in the BGG category ${\cal O}$ of $sl(2)$ representations, we have found a geometric
system whose path integral quantization gives rise to that module.

\section{Illustrations}

In this section we discuss a few instances where the above representations appear in string theory.
In subsection \ref{branesandquantization} we review a connection to the quantization of open
strings ending on branes in a topological A-model, and discuss the representations that
can be obtained in this context. In subsection \ref{coh} we discuss a particle
model with reparameterization invariance that gives rise to a physical cohomology with features
reminiscent of Berkovits cohomology on supercoset sigma-models.

\subsection{Comparison to branes and quantization}
\label{branesandquantization}
In \cite{Gukov:2008ve}
a systematic approach to quantization was proposed using the topological A-model on
a complexification $Y$ of phase space admitting a complete hyper-K\"ahler metric.
 Essential ingredients are the
canonical coisotropic brane ${\cal B}_{cc}$ \cite{Kapustin:2001ij}\cite{Kapustin:2006pk}
with support the whole of $Y$, as well as a Lagrangian A-brane ${\cal B}'$.
The Hilbert or vector space in the quantum theory
consists of the ${\cal B}_{cc}-{\cal B}'$ strings, and the operators acting on them are 
obtained from ${\cal B}_{cc}- {\cal B}_{cc}$ strings which give rise to a non-commutative
algebra of operators. These ideas are explained in detail in  \cite{Gukov:2008ve}, where a comparison
to other approaches to quantization is also made. 

The main illustration of \cite{Gukov:2008ve} is through the A-model on the Eguchi-Hanson
space, viewed as the complexification of the two-sphere. It permits a
natural action of $sl(2)$.  The classical equation for the
two-sphere is translated into a constraint on the $sl(2)$
quadratic Casimir valid in the quantization of the algebra of
polynomials in three variables $h,x,y$. These variables  satisfy the
$sl(2)$ commutation relations in the quantum
theory. Various choices of Lagrangian A-branes, as well as of
parameters in the topological A-model, give rise to a host of
representations of $sl(2)$ on the vector space of ${\cal
  B}_{cc}-{\cal B}'$ strings.

It is shown explicitly in \cite{Gukov:2008ve} how finite dimensional
representations can be associated to a brane wrapping a two-sphere,
and how discrete representations arise from semi-infinite Lagrangian
branes of cigar shape.  It is also discussed how to glue two
representations with semi-infinite spectra into a module with infinite
spectrum, and that the quantization of branes can produce not only
the principal unitary but also 
the complementary series representations (which are otherwise hard to
access via geometric means).

Here, we propose an addendum to the discussion of \cite{Gukov:2008ve}.
Firstly, we remark that our construction of Verma modules $M(\lambda)$
(with positive integer $\lambda$) should correspond to the
quantization of spherical branes glued to semi-infinite Lagrangian
branes, through turning on a vacuum expectation value for a string
between these two branes with one given orientation. The dual Verma
modules arise by turning on a vacuum expectation value for the string
with opposite orientation. Turning on both expectation values would
change the Casimir. This is very close to the discussion of the gluing
of semi-infinite representations in \cite{Gukov:2008ve}.  Secondly, we
remark that by taking the product of two $T^\ast S^2$ spaces, and a
diagonal algebra of ${\cal B}_{cc}-{\cal B}_{cc}$ strings, we are also
able to produce the projective representations $P(-n)$ of $sl(2)$ from
branes and their quantization, through the tensor product
construction. Furthermore, in this way, we also have access to
finite dimensional representations tensored with representations with
infinite spectrum. This considerably extends the class of
representations discussed both in \cite{Gukov:2008ve} as well as in
the bulk of this paper. All of them are geometrically accessible.

\subsection{A reparameterization invariant particle model and its cohomology}
\label{coh}
Consider a particle living on the product of a
hyperboloid (associated to a Verma module) and a sphere (associated to
a finite dimensional representation).  We define the action to
be proportional to the total $sl(2)$ quadratic Casimir invariant $C_2$
(obtained for instance by taking the sums of charges appearing in the
bulk of the paper, and forming the quadratic invariant).  We can also add a mass term
to the action. Thus, we can consider a reparameterization
invariant action of the type $S = \int d \tau e^{-1} ( C_2 - m^2)$ (where
$e$ is an einbein on the worldline of the particle).
When we gauge fix reparameterization invariance (choosing $e=1$), we
add a two-state $(b,c)$ ghost system to the theory.  The BRST operator
is then given by $Q=c(C_2-m^2)$.  In Siegel gauge, we will then have a
cohomology made up of particles satisfying $C_2 = m^2$. Since $C_2^2$
is diagonalizable, we can compute the cohomology in the space of
states satisfying $C_2^2=m^4$.  What happens next depends on the
initial Casimirs and the mass. Either the space $C_2^2=m^4$ has only
one dimensional weight spaces, and the states are in the cohomology. Or,
some of the weight spaces are two-dimensional.  In that case they will
correspond to subspaces of projective representations.  Then we will
have that the operator $C_2-m^2$ maps one state to the other, and the
other to zero. Thus, only the second state will be in the cohomology.
A concrete example is the cohomology of $Q=c \, C_2$ in the space $M(-1)
\otimes L(1) = P(-2)$ (times the two-state ghost system)
which gives rise to a physical state space corresponding to the 
Verma
module $M(0)$. In this example, we took the mass to be zero. Other
examples are easily generated. The resulting cohomologies will
generically project projective representations $P(-n)$ down to highest
weight modules $M(n-2)$. Note that we can define a unitary norm on
this phase space -- it would not be $sl(2)$ invariant.

This construction is a simple analogue of the
calculation of the cohomology for a reparameterization invariant
particle on the supergroup $GL(1|1)$ \cite{Troost:2011fd}. The
calculation is closely related due to the similarity in the action of the quadratic
Casimir on the representation space.

\subsection*{A remark on the literature} 
In \cite{vanTonder:2002gh} a cohomology is defined with respect to the
Casimir operator (minus the mass squared) itself. For instance, it is
found that the cohomology of the quadratic Casimir on the space $M(-1)
\otimes L(1)$ is the space $ L(0)$. Indeed, the doubly degenerate
weight spaces contain one exact and one non-closed state. Neither one
is in the cohomology. One is left with the weight spaces of dimension
one. The construction of \cite{vanTonder:2002gh} is thus quite different from the
one we described above and from the calculation of the physical cohomology in certain
reparameterization invariant string models \cite{Troost:2011fd}.

\section{Conclusions and suggestions}
We provided a path integral description of generalized spin. Geometric models were
proposed that after quantization give rise to all indecomposable representations
in the BGG category ${\cal O}$ of $sl(2)$ modules.  In order to do so, we had to propose
new models for Verma modules that do not belong to the unitary discrete series of $SL(2,\mathbb{R})$,
and provide the path integral realization of the generators of the $sl(2)$ algebra. To realize the
projective representations we used product geometric spaces. 


It would be very interesting to supply elementary models for all the brane quantizations described at the end of
subsection \ref{branesandquantization}, by explicitly 
combining the ingredients of this paper with those of \cite{Gukov:2008ve}. It is then possible to 
give path integral realizations for all simple modules of $sl(2)$ (beyond category ${\cal O}$).
Also, although the category ${\cal O}$ is considerably more complicated for  algebras (and Casimirs) of higher rank, we are
convinced that many of our ideas generalize to a large subset of the representations in those categories.

\section*{Acknowledgements}
It is a pleasure to thank my colleagues
and in particular Samuel Monnier and Catharina Stroppel for interesting discussions.
My work was supported in part by the grant
ANR-09-BLAN-0157-02.

\appendix

\section{Descriptions}
\label{detrep}
In this appendix we collect some 
useful details on $sl(2)$ modules in the BGG category
${\cal O}$. See \cite{HumphreysBGG} for further discussion.
\subsection{The finite dimensional representation $L(\lambda)$}
The finite dimensional modules $L(\lambda)$ with $\lambda$
a positive integer have dimension $\lambda+1$. We can choose
basis vectors $v_{i=0,1,\dots,\lambda}$ on which the algebra
acts as:
\begin{eqnarray}
h \cdot v_i &=&  (\lambda -2 i) v_{i}
\nonumber \\
x \cdot v_i &=& (\lambda - i +1 ) v_{i-1}
\nonumber \\
y \cdot v_i &=& (i+1) v_{i+1}
\nonumber \\
x \cdot v_0 &=& 0
\nonumber \\
y \cdot v_{\lambda} &=& 0.
\end{eqnarray}
The Casimir $C_2=\frac{1}{4}(h^2 + 2h + 4yx)$ is a constant equal to
$C_2 = \frac{\lambda}{4} (\lambda+2)$. In standard physics notation the
spin $j$ of this representation is equal to $j=\lambda/2$
and the Casimir is then $C_2=j (j+1)$.

\subsection{The Verma module $M(\lambda)$}
The Verma module $M(\lambda)$ can be defined as the
module generated by the universal enveloping algebra
acting on a highest weight vector with highest weight
$\lambda$.
The weights of the Verma module
$M(\lambda)$ are $\lambda, \lambda -2, \dots$.
The explicit action of the generators on a basis
consisting of vectors
$v_{0,1,\dots}$ is given by the formulas:
\begin{eqnarray}
h \cdot v_i &=& (\lambda - 2i) v_i
\nonumber \\
x \cdot v_i &=& (\lambda - i + 1) v_{i-1}
\nonumber \\
y \cdot v_i &=& (i+1)  v_{i+1}
\nonumber \\
x \cdot v_0 &=& 0.
\end{eqnarray}
Note that when $\lambda$ is a positive integer,
there will be another maximal vector (besides the vector $v_0$)
inside the Verma module,
namely the vector  $v_{\lambda+1}$. 
 In that case, the maximal
submodule of the module $M(\lambda)$ is $M(-\lambda-2)$. The unique 
simple quotient of the module $M(\lambda)$ is then $L(\lambda)$. When $\lambda$ is not a positive
integer the Verma module $M(\lambda)$ is simple. When the parameter
$\lambda$ is
negative, the representation can be viewed as a discrete
highest weight representation of $sl(2,\mathbb{R})$, often denoted
as $D_{-\frac{\lambda}{2}}^-$, with highest weight $\lambda/2$ and
Casimir $C_2 = j(j-1)$ where $j=-\lambda/2$.

\subsection{The dual Verma module $M(\lambda)^{\vee}$}
Duality acting within the category ${\cal O}$ maps the Verma module
$M(\lambda)$ to its dual $M(\lambda)^\vee$.  The action of the algebra
$sl(2)$ on basis vectors of the module $M(\lambda)^{\vee}$ can be
computed via duality. First of all, duality preserves the weight spaces.  We
can therefore define a dual basis which consists
of maps $w^{j=0,1,\dots}$ such that $w^j(v_i) = \delta^j_i$. Next, we wish
to compute the action of the algebra on this dual vector space.
The anti-involution $\tau$ used to define duality acts as: $\tau(x) =y, \tau(y)=x,
\tau(h) = h$.  The action of the generators on the dual basis
then follows after a small calculation:
\begin{eqnarray}
h \cdot w^j &=& (\lambda-2j) w^j
\nonumber \\
x \cdot w^j &=& j w^{j-1}
\nonumber \\
y \cdot w^j &=&  (\lambda-j) w^{j+1}.
\end{eqnarray}
Note that when 
$\lambda \in \mathbb{N}$,
we have that the basis vector
$w^\lambda$ is annihilated by the {\em lowering} operator $y$.
In that case, there is a non-trivial maximal submodule
which is a $L(\lambda)$.  If we quotient the dual Verma module
$M^\vee(\lambda)$ by the finite dimensional module $L(\lambda)$, we
obtain a module equivalent to the Verma module $M(-\lambda-2)$.  This
illustrates the fact that duality inverts short exact sequences.

\subsection{The projective module $P(-\lambda-2)$}
The projective module $P(-\lambda-2)$ for $\lambda$
a positive integer has weights
$\lambda, \lambda -2, \dots, -\lambda$ with multiplicity one,
and the weights $-\lambda -2, -\lambda -4, \dots$
with multiplicity two. We can choose the action
on the weight vectors $v_{i=0,1,\dots}$ and
$\tilde{v}_{i=0,1,\dots}$ to be:
\begin{eqnarray}
h \cdot v_i &=& (\lambda - 2i) v_i
\nonumber \\
x \cdot v_i &=& (\lambda - i + 1) v_{i-1}
\nonumber \\
y \cdot v_i &=& (i+1)  v_{i+1}
\nonumber \\
h \cdot \tilde{v}_i &=& (-\lambda-2 - 2i) \tilde{v}_i
\nonumber \\
x \cdot \tilde{v}_i  &=& (-\lambda-2 - i + 1) \tilde{v}_{i-1}
+
  \frac{(\lambda+i)!}{i!} v_{\lambda+i} \nonumber \\
y \cdot \tilde{v}_i  &=& (i+1)  \tilde{v}_{i+1}.
\end{eqnarray}
We have a Verma submodule $M(\lambda)$,  as well as
a Verma submodule $M(-\lambda-2)$. It is interesting to compute
the action of the quadratic Casimir on the projective representation
$P(-\lambda-2)$. 
The action on the vectors $v_i$ will be diagonal and equal to
the constant
$\frac{\lambda}{4}(\lambda + 2 )$. The action on the vectors $\tilde{v}_i$ however
is equal to:
\begin{eqnarray}
C_2 \cdot \tilde{v}_i &=& \frac{\lambda}{4}(\lambda+2) \tilde{v}_i
+  \frac{(\lambda+i+1)!}{i!} v_{\lambda + 1 +i}.
\end{eqnarray}
This is neither diagonal nor diagonalizable. If we act with the operator
$(C_2-\frac{\lambda}{4}(\lambda + 2 ))^2$ we will find zero.  The
quadratic Casimir has two-by-two Jordan block structure in the eigenspaces with eigenvalues
$-\lambda-2,-\lambda-4, \dots$

\section{Decompositions}
\label{ETP}
Let's analyze the action of the quadratic Casimir in the tensor
product space $M(\lambda)
\otimes L(1)$:
\begin{eqnarray}
4 C_2  (v_i \otimes \tilde{v}_j) &=&
(h^2+2h + 4 yx) (v_i \otimes \tilde{v}_j)
 \nonumber \\
&=& 
(\lambda^2+2 \lambda+3) (v_i \otimes \tilde{v}_j)
+ 2 (h v_i \otimes h \tilde{v}_j) 
+ 4 (x v_i \otimes y \tilde{v}_j)
+ 4 ( y v_i \otimes x \tilde{v}_j). \nonumber
\end{eqnarray}
One of the last two terms is necessarily zero. Indeed, we have the
actions $x \tilde{v}_0 = 0$ and $y \tilde{v}_1 =0$. On the maximal
vector $v_0 \otimes \tilde{v}_0$ the Casimir acts diagonally with
factor $(\lambda+1)(\lambda+3)/4$.  Now we study the level
one below. It contains a maximal vector (by the fact that the
level above is one-dimensional). Let's compute the $2 \times 2 $
matrix $A$ of the action of the Casimir $4 C_2$ on the vectors $v_1
\otimes \tilde{v}_0$ and $v_0 \otimes \tilde{v}_1$.  We obtain:
\begin{eqnarray}
4 C_2  (v_1 \otimes \tilde{v}_0) &=& 
(\lambda^2+4 \lambda-1) (v_1 \otimes \tilde{v}_0)
+ 4 \lambda ( v_0 \otimes  \tilde{v}_1)
\nonumber \\
4 C_2  (v_0 \otimes \tilde{v}_1) &=& 
(\lambda^2+3) (v_0 \otimes \tilde{v}_1)
+ 4  ( v_1 \otimes  \tilde{v}_0)
\end{eqnarray}
\begin{eqnarray}
A & = & \left(
\begin{array}{cc}
\lambda^2 + 4 \lambda-1 & 4 \lambda \\
4 & \lambda^2+3
\end{array}
\right).
\end{eqnarray}
The matrix $A$ has a characteristic polynomial with zeros
$\lambda^2-1,\lambda^2+4 \lambda+3$, which are the generalized
eigenvalues of the quadratic Casimir. When the eigenvalues are not
equal, the matrix is diagonalizable. Due to the structure of the
weight space, we then identify two direct summands in the tensor
product which are the Verma modules
$M(\lambda+1)$ and $M(\lambda-1)$. On the other hand,
the eigenvalues
are equal when $\lambda=-1$. In this case, the action of the quadratic
central element becomes:
\begin{eqnarray}
\left(
\begin{array}{cc}
-4 & -4 \\
4 & 4
\end{array}
\right)
\end{eqnarray}
which is equivalent to:
\begin{eqnarray}
\left(
\begin{array}{cc}
0 & 0 \\
4 & 0
\end{array}
\right).
\end{eqnarray}
The fact that the action is not diagonalizable demonstrates
that we deal with a projective module, which due to the structure
of the weight spaces must be $P(-2)$. This can be seen a little
more explicitly by identifying $v_0 \otimes \tilde{v}_0$ and
$v_0 \otimes \tilde{v}_1 + v_1 \otimes \tilde{v}_0$ as two highest
weight vectors, and noting that the top state can be reached from
the a non-highest weight vector at the level below. This illustrates the indecomposable 
nature of the module, and its non-trivial
standard filtration in terms of modules $M(0)$ and
$M(-2)$.
As a side remark, we note the similarity between the action of the
raising and lowering operators on the weight spaces of weight $0$ and
$-2$, and the action of fermionic generators within projective
representations of $GL(1|1)$ (see e.g. \cite{Schomerus:2005bf}). This
similarity in the end leads to an isomorphic action of the quadratic
Casimir on weight spaces. (See also subsection \ref{coh}.)


\begin{thebibliography}{99}

\bibitem{Nielsen:1987sa}
  H.~Nielsen and D.~Rohrlich,
  ``A Path Integral To Quantize Spin,''
  Nucl.\ Phys.\ B {\bf 299} (1988) 471.
\bibitem{Johnson:1988qm}
  K.~Johnson,
  ``Functional Integrals For Spin,''
  Annals Phys.\  {\bf 192} (1989) 104.
\bibitem{Alekseev:1988vx}
  A.~Alekseev, L.~Faddeev and S.~Shatashvili,
  ``Quantization of symplectic orbits of compact Lie groups by means of the functional integral,''
  J.\ Geom.\ Phys.\  {\bf 5} (1988) 391.


\bibitem{Ko} B.~Kostant, Quantization and
unitary representations, Modern Analysis and
Applications, Lecture Notes in Math., Vol. 170,
pp. 87-207. Berlin-Heidelberg-New York, Springer 1970.
\bibitem{S} J.~Souriau,
  Structure des syst\`emes dynamiques, Paris, Dunod, 1970.
\bibitem{W} 
  N.~Woodhouse, Geometric quantization, 
New York, Clarendon, 1992.
\bibitem{Ki} A.~Kirillov,
Lectures on the orbit method, Graduate Studies in Mathematics 64, 
American Mathematical Society, Providence, RI, 2004.


\bibitem{HumphreysBGG}
J.~Humphreys,
Representations of Semisimple Lie Algebras in the BGG Category O,  
Grad. Stud. Math., 94, Amer. Math. Soc., Providence, RI, 2008.

\bibitem{Satoh:2001bi}
  Y.~Satoh,
  ``Three point functions and operator product expansion in the SL(2) conformal field theory,''
  Nucl.\ Phys.\ B {\bf 629} (2002) 188
  [hep-th/0109059].

\bibitem{Maldacena:2001km}
  J.~Maldacena and H.~Ooguri,
  ``Strings in AdS(3) and the SL(2,R) WZW model. Part 3. Correlation functions,''
  Phys.\ Rev.\ D {\bf 65} (2002) 106006
  [hep-th/0111180].

\bibitem{Gukov:2008ve}
  S.~Gukov and E.~Witten,
  ``Branes and Quantization,''
  arXiv:0809.0305 [hep-th].

\bibitem{Gotz:2006qp}
  G.~Gotz, T.~Quella and V.~Schomerus,
  ``The WZNW model on $PSU(1,1 \mid 2)$,''
  JHEP {\bf 0703} (2007) 003
  [hep-th/0610070].

\bibitem{Troost:2011fd}
  J.~Troost,
  ``Massless particles on supergroups and $AdS_3 \times S^3$ supergravity,''
  JHEP {\bf 1107} (2011) 042
  [arXiv:1102.0153 [hep-th]].

\bibitem{Gaberdiel:2011vf}
  M.~R.~Gaberdiel and S.~Gerigk,
  ``The massless string spectrum on $AdS_3 \times S^3$ from the supergroup,''
  JHEP {\bf 1110} (2011) 045
  [arXiv:1107.2660 [hep-th]].

\bibitem{Troost:2003ge}
  J.~Troost and A.~Tsuchiya,
  ``Three-dimensional black hole entropy,''
  JHEP {\bf 0306} (2003) 029
  [hep-th/0304211].

\bibitem{Witten:1987ty}
  E.~Witten,
  ``Coadjoint Orbits of the Virasoro Group,''
  Commun.\ Math.\ Phys.\  {\bf 114} (1988) 1.

\bibitem{KoTens}
B.~Kostant, ``On the tensor product of a finite and an infinite
dimensional  representation'',
Journal of Functional Analysis 20, 4 (1975), 257-285.
\bibitem{Z}
G.~Zuckerman, ``Tensor products of finite and infinite dimensional
representations of semisimple Lie groups'',
Annals of Mathematics 106 (1977), 295-308.
\bibitem{BG}
J.~Bernstein and S.~Gelfand, ``Tensor products of
finite and infinite dimensional representations of semisimple
Lie algebras'', 
Compositio Mathematica 41, 2 (1980), 245-285.

\bibitem{BSIII}
J.~Brundan and C.~Stroppel,
``Highest weight categories arising from Khovanov's diagram algebra III: category O'',
 Represent. Theory {\bf 15} (2011), 170-243


\bibitem{Kapustin:2001ij}
  A.~Kapustin and D.~Orlov,
  ``Remarks on A branes, mirror symmetry, and the Fukaya category,''
  J.\ Geom.\ Phys.\  {\bf 48} (2003) 84
  [hep-th/0109098].
\bibitem{Kapustin:2006pk}
  A.~Kapustin and E.~Witten,
  ``Electric-Magnetic Duality And The Geometric Langlands Program,''
  hep-th/0604151.

\bibitem{vanTonder:2002gh}
  A.~van Tonder,
  ``Cohomology and decomposition of tensor product representations of SL(2,R),''
  Nucl.\ Phys.\ B {\bf 677} (2004) 614
  [hep-th/0212149].

\bibitem{Schomerus:2005bf}
  V.~Schomerus and H.~Saleur,
  ``The $GL(1 \mid 1)$ WZW model: From supergeometry to logarithmic CFT,''
  Nucl.\ Phys.\ B {\bf 734} (2006) 221
  [hep-th/0510032].


\end{thebibliography}
\end{document}